\documentstyle[11pt]{article}
\newcommand{\beq}{\begin{equation}}
\newcommand{\eeq}{\end{equation}}

\newcommand{\beqs}{\begin{eqnarray}}
\newcommand{\eeqs}{\end{eqnarray}}

\newcommand{\nn}{\nonumber}

\newcommand{\E}{\cite{E}}

\begin{document}

\begin{center}
\vskip 2.5cm
{\LARGE \bf Is Lavelle-McMullan transformation a really new symmetry in QED?}

\vskip 1.0cm
{\Large  D.~K.~Park }
\\
{\large  Department of Physics,  KyungNam University, Masan, 631-701
Korea}
\vskip 0.5cm

{\Large Hung Soo Kim and Jae Kwan Kim}
\\
{\large Department of Physics, KAIST, Taejon, 305-701, Korea}

\vskip 0.4cm
\end{center}

\centerline{\bf Abstract}

Lavelle-McMullan symmetry of QED is examined at classical and quantum levels.
It is shown that Lavelle-McMullan symmetry does not give any new non-trivial
information in QED by examining the Ward-Takahashi identities. Being inspired
by the examination of Ward-Takahashi identity, we construct the generalized
non-local and non-covariant symmetries of QED.

\vfill

\newpage
\setcounter{footnote}{1}

In a recent work Lavelle-McMullan(LM) found a non-local and non-covariant
symmetry of QED in the Feynman gauge[1]. They claimed that this is new
symmetry and the physical states of QED are restricted by the conditions
\beqs
Q_{BRST} \mid phys >&=& 0,   \\ \nn
  Q_{LM} \mid phys >&=& 0,
\eeqs
where $Q_{BRST}$ and $Q_{LM}$ are charges of BRST[2, 3] and LM transformations
respectively. More recently it was disputed whether LM symmetry is merely
non-local version of BRST symmetry or not[4, 5].

In the present paper we will examine this subject thoroughly by deriving
the Ward-Takahashi(WT) identities and show that LM symmetry does
not yield any new
non-trivial information. In addition we will construct the generalized
non-local and non-covariant symmetry for QED.

The QED Lagrangian we will study in Feynman gauge is
\beq
{\cal L} = - \frac{1}{4}F_{\mu \nu}F^{\mu \nu} -
          \frac{1}{2}(\partial_{\mu}A^{\mu})^2 +
          \bar{\psi}(i \gamma^{\mu}D_{\mu} - m) \psi -
          i \partial_{\mu} \bar{c} \partial^{\mu} c,
\eeq
where $D_{\mu} = \partial_{\mu} + ig A_{\mu}$.
The BRST symmetry for the Lagrangian (2)
\beqs
    \delta_{BRST}A_{\mu}&=& \partial_{\mu}c,  \\ \nn
          \delta_{BRST}c&=& 0, \\ \nn
   \delta_{BRST} \bar{c}&=& -i \partial_{\mu}A^{\mu},  \\ \nn
      \delta_{BRST} \psi&=& -igc \psi,\\ \nn
\delta_{BRST} \bar{\psi}&=& -ig \bar{\psi} c,
\eeqs
yields a conserved current through Noether's theorem
\beq
J^{\mu}_{BRST} = -(F^{\mu \nu} + g^{\mu \nu} \partial_{\rho}A^{\rho})
                 \partial_{\nu} c + g J^{\mu}c,
\eeq
where $J^{\mu} = \bar{\psi} \gamma^{\mu} \psi$.

On the other hand LM symmetry
\beqs
\delta_{LM}A_0&=& i \bar{c}, \\ \nn
\delta_{LM}A_i&=& i \frac{\partial_i \partial_0}{\nabla^2} \bar{c}, \\ \nn
\delta_{LM}c&=& A_0 - \frac{\partial_i \partial_0}{\nabla^2} A_i +
               \frac{g}{\nabla^2} J_0,  \\ \nn
\delta_{LM} \bar{c}&=& 0,  \\ \nn
\delta_{LM} \psi&=& (\frac{g}{\nabla^2} \partial_0 \bar{c}) \psi, \\ \nn
\delta_{LM} \bar{\psi}&=& \bar{\psi} ( \frac{g}{\nabla^2} \partial_0 \bar{c}),
\eeqs
yields another conserved current
\beqs
J^{\mu}_{LM} =&-&i(F^{\mu 0} + g^{\mu 0} \partial_{\rho}A^{\rho}) \bar{c}
               -i(F^{\mu i} + g^{\mu i} \partial_{\rho}A^{\rho})
               \frac{\partial_i \partial_0}{\nabla^2} \bar{c}  \\ \nn
              &+&i J^{\mu}(\frac{g}{\nabla^2} \partial_0 \bar{c})
               + i \Box \bar{c} \frac{F^{0 \mu}}{\nabla^2}.
\eeqs
Using equations of motion
\beqs
g J^{\nu}&=& \partial_{\mu}(F^{\mu \nu} + g^{\mu \nu} \partial_{\rho}A^{\rho}),
\\  \nn
(i \gamma^{\mu} D_{\mu} - m) \psi&=& 0, \\ \nn
\Box c = \Box \bar{c}&=& 0,
\eeqs
one can straightforwardly prove
\beq
\partial_{\mu}J^{\mu}_{LM} = i c^{-1}(\frac{\partial_0}{\nabla^2} \bar{c})
                     \partial_{\mu}J^{\mu}_{BRST}.
\eeq
Eq.(8) implys that BRST symmetry guarantees the symmetric property of LM
transformation even at the classical level. This means LM symmetry is not
independent of BRST symmetry.

In order to study the quantum theory  let us examine
the WT identities. Taking a LM variation to the well-known identities of
QED
\beq
<0 \mid T A_{\mu}(x)c(y) \mid 0> = 0,
\eeq
one gets
\beqs
-i D(q) + D_{00}(q) - \frac{q_i q_0}{\vec{q}^2} D_{0i}(q) -
\frac{g}{\vec{q}^2} \Lambda_{00}(q) = 0,    \\  \nonumber
-i \frac{q_i q_0}{\vec{q}^2} D(q) + D_{i0}(q) -
\frac{q_j q_0}{\vec{q}^2} D_{ij}(q) - \frac{g}{\vec{q}^2}
\Lambda_{i0}(q) = 0,
\eeqs
where
\beqs
D(q)&=& \int dx e^{iq\cdot x} <0 \mid T c(x) \bar{c}(0) \mid 0>,  \\  \nonumber
D_{\mu \nu}(q)&=& \int dx e^{iq \cdot x} <0 \mid T A_{\mu}(x) A_{\nu}(0) \mid
0>,
 \\  \nonumber
\Lambda_{\mu \nu}(q)&=& \int dx e^{iq \cdot x} <0 \mid T A_{\mu}(x) J_{\nu}(0)
\mid 0>.
\eeqs
If one divides $D_{\mu \nu}(q)$ as
\beq
D_{\mu \nu}(q) = D^{(0)}_{\mu \nu}(q) + \tilde{D}_{\mu \nu}(q),
\eeq
where $ D^{(0)}_{\mu \nu}(q) $ is a propagator of gauge field at the tree
level which can be obtained from usual Feynman rule and $\tilde{D}_{\mu
\nu}(q)$
 is a higher order correction, Eq.(10) becomes for n-loop order
\beqs
-i D(q) + D^{(0)}_{00}(q)&=& 0 \hspace{.5cm} for \hspace{.5cm} n=0,  \\ \nn
-i \frac{q_i q_0}{\vec{q}^2} D(q) - \frac{q_j q_0}{\vec{q}^2} D^{(0)}_{ij}(q)
&=& 0   \hspace{.5cm}  for \hspace{.5cm} n=0,  \\ \nn,
\tilde{D}_{00}(q) - \frac{q_i q_0}{\vec{q}^2} \tilde{D}_{0i}(q) -
\frac{g}{\vec{q}^2} \Lambda_{00}(q)&=& 0  \hspace{.5cm} for \hspace{.5cm} n
\geq 1, \\ \nn
\tilde{D}_{i0}(q) - \frac{q_j q_0}{\vec{q}^2} \tilde{D}_{ij}(q) -
\frac{g}{\vec{q}^2} \Lambda_{i0}(q)&=& 0 \hspace{.5cm} for \hspace{.5cm} n \geq
1.
\eeqs
One can show easily that the first two equations of Eq.(13) become
\beq
q^{\mu}D^{(0)}_{\mu \nu}(q) = i q_{\nu} D(q).
\eeq
By using the trivial identity
\beq
\Lambda_{\mu \nu}(q) = - \frac{q^2}{g} \tilde{D}_{\mu \nu}(q)
\eeq
which is valid in n($\geq 1$)-loop level and can be easily proved
by Feynman diagram, the last two equations of Eq.(13) become
\beq
q^{\mu} \tilde{D}_{\mu \nu} = 0.
\eeq
Fig.(1) shows the example of trivial identity (15) at one-loop level.
Combining (14) and (16) one can obtain WT identity
\beq
q^{\mu} D_{\mu \nu}(q) = i q_{\nu} D(q).
\eeq
Eq.(17) is well-known WT identity which is derived from the usual BRST
symmetry. So in this case LM symmetry does not produce new WT identity.
Now let us examine the WT identity which contains the vertex. For this
let us start with a identity of QED
\beq
<0 \mid T \psi(x) \bar{\psi}(y) c(z)\mid 0> = 0.
\eeq
By following the same procedure LM symmetry provides
\beqs
ig \frac{q_0}{\vec{q}^2}D(q)[S_F(p+q) - S_F(p)]
+ \Gamma_0(p, q, -p-q) -        \\  \nonumber
\frac{q_i q_0}{\vec{q}^2} \Gamma_i(p, q, -p-q)
- \frac{g}{\vec{q}^2}K_0(p, q, -p-q) = 0
\eeqs
where
\begin{eqnarray}
S_F(q)&=& \int dx e^{iq \cdot x} <0 \mid T \psi(x) \bar{\psi}(0) \mid 0>,  \\
\nn
<0 \mid T \psi(x) A_{\mu}(z) \bar{\psi}(y) \mid 0>&=& \int dp dq dr
\Gamma_{\mu}
(p, q, r) e^{-ip \cdot x -iq \cdot z -ir \cdot y} \delta(p+q+r),   \\
\nonumber
<0 \mid T \psi(x) J_{\mu}(z) \bar{\psi}(y) \mid 0>&=& \int dp dq dr
K_{\mu}(p, q, r) e^{-ip \cdot x -iq \cdot z -ir \cdot y} \delta(p+q+r).
\end{eqnarray}
In Eq.(20) an energy-momentum conserving delta function is included in the
vertex-type objects. From Fig.(2) one can prove straightforwardly the trivial
identity
\beq
K_{\mu}(p, q, -p-q) = - \frac{q^2}{g} \Gamma_{\mu}(p, q, -p-q).
\eeq
By using Eq.(21) the WT identity
\beq
q^{\mu} \Gamma_{\mu}(p, q, -p-q) = -ig D(q) [S_F(p+q) - S_F(p)]
\eeq
is straightforwardly derived from Eq.(19).
The WT identity (22) can be easily re-derived
from the BRST symmetry and gives rise to a famous formula
\beq
Z_1 = Z_2
\eeq
by the Dyson Z-notation.So LM symmetry does not give any new non-trivial
identity in this
case too.

Being inspired by the examination of WT identities, we can construct
the generalized
symmetries of QED
\beqs
\delta_{G}A_0&=& i \bar{c},  \\  \nn
\delta_{G}A_i&=& -\partial_i a(-i \partial) \bar{c},  \\ \nn
\delta_{G}\bar{c}&=& 0,  \\  \nn
\delta_{G}c&=& A_0 -i \partial_i a(-i \partial) A_i + b(-i \partial) J_0
              -i \partial_i f(-i \partial) J_i,   \\  \nn
\delta_{G}\psi&=& (g(-i \partial) \bar{c}) \psi,   \\ \nn
\delta_{G}\bar{\psi}&=& \bar{\psi} ( g(-i \partial) \bar{c}),
\eeqs
where $a(x)$, $b(x)$, $f(x)$ and $g(x)$ are functions of four vector.
If one requires the conditions
\beqs
x_0a(x) - \frac{x^2}{g}[x_0 f(x) + b(x)]&=& -1,  \\ \nn
g(x)&=& ig[a(x) - \frac{x^2}{g} f(x)],  \\ \nn
a(-x)&=& - a(x),  \\ \nn
b(-x)&=& b(x),   \\ \nn
f(-x)&=& -f(x),
\eeqs
it is straightforward to prove that the transformation(24) with restrictions
(25) is symmetry for the QED Lagrangian(2). Of course LM transformation is
a special case of (24):
\beqs
a_{LM}(x)&=& -\frac{x_0}{\vec{x}^2},   \\  \nn
b_{LM}(x)&=& -\frac{g}{\vec{x}^2},  \\  \nn
f_{LM}(x)&=& 0,  \\  \nn
g_{LM}(x)&=& -ig \frac{x_0}{\vec{x}^2}.
\eeqs
By the same procedure one can show easily that this generalized symmetry(24)
with (25) provides same WT identities (17) and (22) using the trivial
identities(15) and (21).

In conclusion although LM transformation is non-local and non-covariant
symmetry, not only LM symmetry but also the generalized symmetry does not
yield any new non-trivial identity at quantum level.

\bigskip



\newpage
\input FEYNMAN
\begin{figure}
\begin{picture}(20000,10000)
\thicklines
\put(-6000,4000){$A_{\mu}$}
\put(9500,4000){$A_{\nu}$}
\put(14500,4000){$A_{\mu}$}
\put(26000,4000){$J_{\rho}$}
\put(31000,4000){$A^{\rho}$}
\put(38500,4000){$A_{\nu}$}
\drawline\gluon[\E\REG](-4000,4000)[4]
\put(2500,4000){\circle{4000}}
\drawline\gluon[\E\REG](4500,4000)[4]
\put(12500,4000){=}
\drawline\gluon[\E\REG](16500,4000)[4]
\put(23000,4000){\circle{4000}}
\put(29000,4000){X}
\drawline\gluon[\E\REG](33000,4000)[4]
\end{picture}

\caption{one-loop example of trivial identity(15)}
\label{Figure 1}

\end{figure}
\pagebreak
\begin{figure}
\begin{picture}(20000,10000)
\thicklines
\drawline\gluon[\E\REG](-4000,4000)[4]
\drawline\fermion[\NW\REG](\particlefrontx,\particlefronty)[4500]
\drawline\fermion[\SW\REG](\particlefrontx,\particlefronty)[2500]
\global\Xone=\particlebackx
\global\Yone=\particlebacky
\drawline\fermion[\SW\REG](\Xone,\Yone)[2000]
\drawline\gluon[\N\REG](\Xone,\Yone)[3]

\put(2500,4000){\circle{4000}}
\drawline\gluon[\E\REG](4500,4000)[4]
\put(9000,4000){$A_{\mu}$}
\put(11000,4000){=}
\drawline\gluon[\E\REG](18000,4000)[4]
\drawline\fermion[\NW\REG](\particlefrontx,\particlefronty)[4500]
\drawline\fermion[\SW\REG](\particlefrontx,\particlefronty)[2500]
\global\Xone=\particlebackx
\global\Yone=\particlebacky
\drawline\fermion[\SW\REG](\Xone,\Yone)[2000]
\drawline\gluon[\N\REG](\Xone,\Yone)[3]

\put(24500,4000){\circle{4000}}
\put(27500,4000){$J_{\rho}$}
\put(30500,4000){X}
\drawline\gluon[\E\REG](34500,4000)[4]
\put(40000,4000){$A_{\mu}$}
\put(32500,4000){$A^{\rho}$}
\end{picture}

\caption{two-loop example of trivial identity(21)}
\label{Figure 2}

\end{figure}

\end{document}